\documentclass[11pt]{article}
\usepackage{latexsym,amssymb,graphicx}

 \def\pd{\partial} \def\pp{\prime}  \def\b{\beta} \def\dl{\delta} \def\s{\sigma}  \def\eps{\epsilon}  \def\lam{\lambda} \def\Lam{\Lambda}     \def\sq{\sqrt} \def\fr{\frac} 
 
 \def\hg{{\hat g}} \def\bg{{\bar g}}  
  \def\bnb{{\bar \nabla}}  
\def\bDelta{{\bar \Delta}} 

\def\bx{{\bf x}}  
\def\lap3{~| \!\!\! \partial^2} \def\dlap3{~| \!\!\! \partial^4} \def\invlap3{~| \!\!\! \partial^{-2}}

\begin{document}

\begin{center}
{\Large {\bf Localized massive excitation of quantum gravity as a dark particle}} 
\end{center}

\begin{center}
{\sc Ken-ji Hamada}
\end{center}

\begin{center}
{\it Institute of Particle and Nuclear Studies, KEK, Tsukuba 305-0801, Japan  \\ and
Department of Particle and Nuclear Physics, The Graduate University for Advanced Studies (SOKENDAI), Tsukuba 305-0801, Japan}
\end{center}

\begin{abstract}
We construct a static and spherical excited state without singularities in renormalizable quantum gravity with background-free nature asymptotically. Its diameter is given by a correlation length of the quantum gravity, longer than the Planck length by 2 orders of magnitude, and it has a Schwarzschild tail outside. The quantum gravity dynamics inside is described by employing a nonperturbative expression of higher-order corrections assumed from a physical requirement that the dynamics disappear at the edge where it is in strong coupling. A running coupling constant that is a manifestation of nonlinearity and nonlocality is managed by approximating it as a mean field that depends on the radial coordinate. If the mass is several times the Planck mass, we can set up a system of linearized equations of motion for the gravitational potentials incorporating the running effect and obtain the excited state as its solution. It may be a candidate for dark matter, and will give a new perspective on black hole physics.
\end{abstract}

\vspace{3mm}

\section{Introduction}

In Einstein's theory of gravity, a pointlike particle with mass beyond the Planck scale is a black hole. In fact, the Compton wavelength of such a particle is shorter than the horizon size of the mass, and thus its information is confined and lost. Hence, describing the particle as a point is no longer justified. This is the reason why the Planck scale has been recognized as a wall that cannot be exceeded. Here, to overcome such a problem, we argue that spacetime will transition to a new phase before reaching the wall, and a quantum world without singularities will come out. Renormalizable asymptotically background-free quantum gravity \cite{hs, hamada02, hamada14, hm16, hm17, book} formulated based on a certain conformal field theory \cite{riegert, am, amm92, amm97, hh, hamada12M4, hamada12RxS3, book} suggests the existence of a dynamical energy scale that clearly separates quantum spacetime from classical spacetime.

The action is given by the sum of a conformally invariant gravity part $I_4 = \int d^4 x \sq{-g} [-C^2_{\mu\nu\lam\s}/t^2 -b G_4]$ and other lower derivative terms $I_L = \hbar^{-1} \int d^4 x \sq {-g} [M^2 R/2 - \lam + \cdots]$, where $C_{\mu\nu\lam\s}$ is the Weyl tensor and $G_4$ is the Euler density. The dots denote matter actions that become conformally invariant in the ultraviolet (UV) limit. For simplicity, bare and renormalized quantities are not distinguished here. $t$ is a dimensionless coupling constant which represents a deviation from conformally flat configurations. The perturbation by $t$ is justified because the beta function becomes negative. $b$ is introduced to remove UV divergences proportional to the Euler term, and is not an independent one, but expanded by $t$. Since the gravitational field is exactly dimensionless, $I_4$ is a dimensionless action, thus has no $\hbar$.

One of the reasons for considering the positive definite action involving the square of  the Riemann curvature tensor is that singularities become unphysical because the action diverges, unlike the Einstein-Hilbert action that is not even bounded below. Nevertheless, simply applying perturbation theory to such a fourth derivative gravity causes the problem of ghosts. In order to solve it, we need to apply a nonperturbative method developed from two-dimensional quantum gravity \cite{polyakov, kpz, dk, david}.

The metric field is decomposed as $g_{\mu\nu}=e^{2\phi}\bg_{\mu\nu}$ and $\bg_{\mu\nu} = (\hg e^{h})_{\mu\nu} = \hg_{\mu\lam} ( \dl^\lam_{~\nu}  + h^\lam_{~\nu} + \cdots )$, where $h^\mu_{~ \nu}$ is the traceless tensor field that is controlled by $t$.\footnote{
Normally, renormalization is done by replacing  $h^\mu_ {~\nu}$ with $th^\mu_{~\nu}$. On the other hand, $\phi$ has no coupling constant, thus the non-renormalization theorem holds for it \cite{hamada02, hamada14, hm16, hm17}.} 
The method is that a diffeomorphism invariant measure $[dg]_g [df]_g$ is rewritten using a practical measure defined on the background metric $\hg_{\mu\nu}$ as $[dg]_g [df]_g = [d\phi]_\hg [dh]_\hg [df]_\hg e^{iS}$, where matter fields are symbolically represented by $f$. $S$ is the Wess-Zumino action for conformal anomalies \cite{cd, ddi, duff}, that is Jacobian to preserve diffeomorphism invariance. In particular, a zeroth-order term of $t$ exists in $S$ and is given by the Riegert action \cite{riegert},  $S_{\rm R} = -(b_c/16\pi^2) \int d^4 x \sq{-\bg} [ 2 \phi {\bar \Delta}_4 \phi + ({\bar G}_4 -2\bnb^2 \bar{R}/3)\phi  ]$, where $\sq{-g} \Delta_4$ is a conformally invariant fourth derivative differential operator. The quantities with the bar are the ones defined by $\bg_{\mu\nu}$. This action gives a kinetic term of the conformal factor field $\phi$. The coefficient $b_c$ has a right sign and is about $10$ in ordinary particle models including various grand unified theories; here, $b_c=10$ is employed.

Writing the whole action as ${\cal I}_{\rm 4DQG} = S + I_4 + I_L$, we can rewrite the path integral to $\int [d\phi \, dh \, df]_\hg e^{i {\cal I}_{\rm 4DQG}}$ as a standard quantum field theory on $\hg_{\mu\nu}$. Here, neither $I_4$ nor $S$ depends on $\hbar$; thus, together they are an action describing quantum gravity states in the world beyond the Planck scale. In the following, $\hbar = 1$.

The most significant feature of the theory is that a conformal invariance remains as part of diffeomorphism invariance at the UV limit of $t \to 0$ \cite{hh, hamada12M4, hamada12RxS3}. Here, to emphasize it is a gauge symmetry, namely, Becchi-Rouet-Stora-Tyutin (BRST) symmetry, we call it BRST conformal symmetry. It means that all theories with different backgrounds connected to each other by conformal transformations are gauge equivalent. Therefore, the flat metric can be employed as a background without affecting the physics. The appearance of this gauge symmetry in the UV limit is called ``asymptotic background freedom.'' Although negative-metric modes are necessary components for the BRST conformal algebra to close at the quantum level, we can show that they all are not gauge invariant, and thus do not appear as physical states.\footnote{
Physical fields are given by real primary scalars only \cite{amm97, hh, hamada12M4, hamada12RxS3, book}. Their reality will be guaranteed by the positivity of the whole action.} 

The theory has three physical scales, namely, renormalization group invariants \cite{hm17} that must be determined experimentally. They are the Planck mass $m_{\rm pl}=1/\sq{G} \simeq 10^{19}$GeV, the cosmological constant that is ignored here, and the dynamical scale $\Lam_{\rm QG}$ originated from the negativity of the beta function.  The last one is the scale that separates quantum and classical spacetimes and is predicted to be $\Lam_{\rm QG} \simeq 10^{17}$GeV from an inflation scenario driven by quantum gravity dynamics only \cite{hy, hhy06, hhy10, book}.

\section{Equations of motion}
Considering spherical scalar fluctuations of gravity and assuming they are small, we write the metric as $ds^2 = -(1+2\Psi) d\eta^2 + (1+ 2\Phi) d\bx^2$. Here, $\Phi$ and $\Psi$ are called the gravitational potentials. The linearized Einstein equation around a static point mass is then expressed as $-2M_{\rm P}^2 \lap3 \Phi =0$ and $\Psi = -\Phi$, where $M_{\rm P}=1/\sq{8\pi G}$ is the reduced Planck mass and $\lap3 = \pd^2_r + (2/r) \pd_r$ is the spatial Laplacian in which the radial derivative is written as $\pd_r = \pd/\pd r$. For small mass $m$, the gravitational potential is given by $\Phi =r_g/2r$, where $r_g = 2Gm$ is the Schwarzschild radius.

Here we consider a localized excited state of the quantum gravity that has a tail of the Schwarzschild solution outside. The diameter of the excited state will be a correlation length given by the inverse of the dynamical scale denoted as $\xi_\Lam =1/\Lam_{\rm QG}$, and thus, the radius is $R_h=\xi_\Lam/2 \simeq 10^{-31}{\rm cm}$. The magnitude of the gravitational potentials will not monotonically increase inside. Hence, if $2\Phi$ is sufficiently smaller than $1$ at the edge, namely, $r_g \ll R_h$, then applying linear approximation for the gravitational potentials will become effective. The condition is expressed as $m \ll m_{\rm pl}^2/4\Lam_{\rm QG}$, and the right-hand side is about $25$ times $m_{\rm pl}$. In the following, the calculations will be performed assuming that the linear approximation holds even near the origin, and the validity will be confirmed from the result. Fortunately, we can find a solution with a mass of about the Planck mass, which is the magnitude we most wanted to know.\footnote{ 
There are two main sources causing nonlinearity, one is from the coupling constant $t$ involving the running effect described later, and the other is due to the exponential factor of $\phi$ contained in the Einstein-Hilbert action that arises even at $t=0$. If the mass is even larger, the latter nonlinearity cannot be ignored near the origin, even though it is small at the edge.} 

The inside of the excited state is expressed by the quantum gravity. The conformal gravity dynamics near the origin is described by the Riegert action and linearized Weyl action, while 
the vicinity of the boundary separating the inside and the outside will be a strong coupling region of $t$; thus, its effect should be involved. It is known that the Riegert action receives loop corrections so that the coefficient is replaced as $b_c \to b_c (1 - a_1 t^2 + \cdots)=b_c B(t)$ \cite{hs, hamada02, hamada14, hm16, hm17, book}. As a physical requirement, the conformal gravity dynamics should disappear in the strong coupling limit. So, here we assume the form boldly summed up as $B(t)=[1+a_1 t^2]^{-1}$ as the simplest nonperturbative expression satisfying such a requirement \cite{hhy06}.

The energy-momentum tensor consists of three terms as $T_{\mu\nu}=T^{(4)}_{\mu\nu} + T^{\rm EH}_{\mu\nu} + T^{\rm M}_{\mu\nu}$. The first is derived from the fourth derivative terms $S+I_4$, the second is from the Einstein-Hilbert action, and the last is from matter fields. The matter part is here represented as a relativistic perfect fluid, and its energy density is denoted by $\rho$.

The equations of motion are given by $T_{\mu\nu}=0$. There are two combinations in which $\rho$ disappears and result in equations for the gravitational potentials only\footnote{
Equations (\ref{trace equation}) and (\ref{constraint equation}) are (4.16) and (4.17) multiplied by $\lap3$ with taking $\phi(\eta) =0$ in \cite{hhy06}, respectively. Equation (\ref{energy conservation}) is from (4.20), in which $\rho D \, (=\dl \rho)$ is rewritten as $\rho$.}:  
\begin{eqnarray}
 \!\!\!\! \!\!\!\! \!\!\!
   && \fr{b_c}{8\pi^2} B(t) \biggl( 
                       \!-\!   2 \pd_\eta^4 \Phi   
                       \!+\!   \fr{10}{3} \pd_\eta^2 \lap3 \Phi  
                       \!-\!   \fr{4}{3} \dlap3 \Phi   
                       \!+\!    \fr{2}{3} \pd_\eta^2 \lap3 \Psi  
               \nonumber \\
 \!\!\!\! \!\!\!\! \!\!\!
   &&                \!-\! \fr{2}{3} \dlap3 \Psi     \biggr)
                       \!+\!    M_{\rm P}^2 \bigl(    6 \pd_\eta^2 \Phi  
                       \!-\!    4 \lap3 \Phi -2 \lap3 \Psi    \bigr) 
                       \!=\!   0 
               \label{trace equation} 
\end{eqnarray}
and
\begin{eqnarray}
 \!\!\!\! \!\!\!\! \!\!\!
    &&  \fr{b_c}{8\pi^2} B(t) \biggl( \fr{4}{3} \pd_\eta^2 \lap3 \Phi   
                    \!-\!   \fr{8}{9} \dlap3 \Phi   
                    \!-\!   \fr{4}{9} \dlap3 \Psi    \biggr)
                    \!+\!   \fr{2}{t^2} \biggl( 4 \pd_\eta^2 \lap3 \Phi  
                     \nonumber \\
 \!\!\!\! \!\!\!\! \!\!\!
   &&             \!-\!   \fr{4}{3} \dlap3 \Phi 
                    \!-\!   4 \pd_\eta^2 \lap3 \Psi   
                    \!+\!   \fr{4}{3} \dlap3 \Psi     \biggr)
                    \!-\!   2 M_{\rm P}^2 \bigl(  \lap3 \Phi   
                    \!+\!   \lap3 \Psi    \bigr)  
                    \!=\!   0 .
             \label{constraint equation}
\end{eqnarray}
The part containing $b_c$ is derived from the Riegert action, the part with $1/t^2$ is from the Weyl action, and $M_{\rm P}^2$ is from the Einstein-Hilbert action. In addition, from the time-time component, an energy conservation equation containing $\rho$ is yielded as
\begin{eqnarray}
   &&  \fr{b_c}{8\pi^2} B(t) \biggl(  - \fr{2}{3} \pd_\eta^2 \lap3 \Phi  
         + \fr{4}{9} \dlap3 \Phi   + \fr{2}{9} \dlap3 \Psi   \biggr)
                \nonumber \\
   && + \fr{2}{t^2} \biggl( - \fr{4}{3} \dlap3 \Phi  + \fr{4}{3} \dlap3 \Psi  \biggr)
        + 2 M_{\rm P}^2 \lap3 \Phi  +  \rho = 0  .
         \label{energy conservation}
\end{eqnarray}

These equations do not yet include the dynamics involving the correlation length $\xi_\Lam$. Including such dynamical effects is not simple, and equations of motion become rather complicated, containing nonlinear as well as nonlocal terms. A running coupling constant representing them is an operator that acts on the field, usually defined by $\bar{t}^2(Q)=[\b_0 \log (Q^2\xi_\Lam^2)]^{-1}$, where $Q$ is a physical momentum and $\b_0$ is a coefficient of the beta function.\footnote{
\label{footnote on running coupling}It is defined through the fact that the effective action for the Weyl sector $- [ \, 1/t^2 - 2\b_0 \phi + \b_0 \log (q^2/\mu^2) \, ] \sq{-g} \, C_{\mu\nu\lam\s}^2$ can be written in the form $- \sq{-g} \, C_{\mu\nu\lam\s}^2/\bar{t}^2 (Q)$ with $Q^2=q^2/e^{2\phi}$ defined in the full metric space, where $q$ is a momentum defined on the background \cite{hamada02, hhy06}. The dynamical scale is $\Lam_{\rm QG} = \mu e^{-1/2\b_0 t^2}$ with $\mu$ being a renormalization mass scale. The $\phi$ term is one of the Wess-Zumino actions for conformal anomaly. The argument is generalized to higher loops; there, interactions of the type $\phi^n C^2_{\mu\nu\lam\s} ~(n \geq 2)$ are involved. This indicates that the Wess-Zumino action appears in the form so that $t^2$ is replaced with $\bar{t}^2(Q)$ due to diffeomorphism invariance.}  
Here, we bravely simplify the equations of motion by representing the running coupling constant as an average under the spirit of the mean field approximation \cite{hhy06}, that is, rewriting it as a coordinate-dependent average by replacing $Q$ with the inverse of $2r$, so that
\begin{eqnarray}
     \bar{t}^2(r) = [\b_0 \log (R_h^2/r^2)]^{-1} ,
         \label{running coupling}
\end{eqnarray}
which diverges at the edge $r=R_h$. Then we replace $t^2$ in (\ref{trace equation}), (\ref{constraint equation}), and (\ref{energy conservation}) with $\bar{t}^2(r)$. At the same time, $B$ is replaced with $\bar{B}(r) = [1+a_1 \bar{t}^2(r) ]^{-1}$.\footnote{
This replacement reflects the appearance of the Wess-Zumino interactions of the type $\phi^{n+1} \bDelta_4 \phi ~( n \geq 1)$ at higher orders in $t$, where $n=0$ is the Riegert action \cite{hamada02,hhy06}.} 

The running coupling constant (\ref{running coupling}) is a function that vanishes at the origin, gradually increases away from there, and diverges sharply at the edge $r=R_h$, which plays the role of connecting the inside and outside of the excited state. Indeed, taking it into account, (\ref{constraint equation}) shows that around the origin with $\bar{t} \simeq 0$, configurations in which the inside of the second parenthesis vanishes become dominant, whereas at the edge where $\bar{t} \to \infty$, the fourth derivative conformal dynamics disappear, leading to Einstein gravity.

In this way, we set up the equations of motion incorporating the strong coupling dynamics near the edge as a function of the radial coordinate while maintaining linearity of the gravitational potentials. The parameters $\b_0$ and $a_1$ are rather vague, unlike the coefficient $b_c$, because they are concerned with the corrections by $t$ and are built into the nonperturbative expression assumed. They should be regarded as free parameters for specifying the strong coupling dynamics here; hence, it is adequate to determine them phenomenologically. Here, $\b_0 = 0.5$ and $a_1 = 0.1$ are employed, which are almost the same as those used in the inflation scenario \cite{hhy06, hhy10}.

\section{Spherical excitations}
Let us consider a static solution of the equations of motion. In this case, by introducing new variables $X=2\Phi + \Psi$ and $Y=\Phi-\Psi$, the coupled equations (\ref{trace equation}) and (\ref{constraint equation}) can be completely divided into two as $\bar{B} \dlap3 X + 3 H_{\rm D}^2 \lap3 X = 0$ and $(4/\bar{t}^2) \dlap3 Y - M_{\rm P}^2 \lap3 Y = 0$, where $H_{\rm D}= M_{\rm P} \sq{8\pi^2/b_c}$, which has a value between the Planck mass and the reduced Planck mass. The static solution also satisfies the energy equation (\ref{energy conservation}) of $\rho = 0$; thus, it represents a purely gravitational excitation.

We further rewrite the variables as $X(r) = (r_g/2 r) f(r)$ and $Y(r) = (r_g/r) g(r)$, where $r_g$ is an unknown constant for the time being. The relationship between $r_g$ and mass given before will be determined from the energy equation (\ref{energy conservation}) later. Applying the spatial Laplacian to these variables gives $\lap3 X = (r_g/2r) \, \pd^2_r f$ and $\dlap3 X = (r_g/2r) \, \pd^4_r f$. The same applies to the variable $Y$. We then obtain
\begin{eqnarray}
     \pd_r^4 f(r) + 3H_{\rm D}^2 [1 + a_1 \bar{t}^2(r) ] \pd_r^2 f(r) &=& 0,
            \nonumber \\
     \pd_r^4 g(r) - \fr{1}{4} M_{\rm P}^2 \bar{t}^2(r) \pd_r^2 g(r) &=& 0 .
        \label{equations of f and g}
\end{eqnarray}

As conditions that the variables $X$ and $Y$ and $\lap3 X$ and $\lap3 Y$ do not diverge at the origin, we obtain
\begin{eqnarray}
      f(0) = 0,   \quad  g(0) = 0, \quad  \pd_r^2 f(0) = 0,   \quad \pd_r^2 g(0) = 0 .
            \label{conditions at origin}
\end{eqnarray}
As stated in the Introduction, singularities are unphysical; thus, solutions in which $f$ or $g$ is finite at the origin are excluded because they are fake ones caused by employing the linear approximation. The conditions given by the second derivative of $r$ are requirements for each term of the energy equation (\ref{energy conservation}) to be finite. Furthermore, as conditions for smoothly  connecting to the Schwarzschild solution at the edge, we set
\begin{eqnarray}
     f(R_h) = 1,   \,\,  g(R_h) = 1, \,\,   \pd_r f(R_h) = 0, \,\,  \pd_r g(R_h) = 0 .
         \label{conditions at edge}
\end{eqnarray}
The first two show that $\Phi(R_h)= -\Psi(R_h)$ holds as the Einstein equation does. The last two conditions are for smooth connection.

First, we examine the behavior of the solution near the origin. Letting $\zeta = \pd_r^2 f$ and $\theta = \pd_r^2 g$, we look for solutions that satisfy $\zeta(0)=\theta(0)=0$. In addition, since the running coupling constant is so small around the origin, it is taken to be a small constant such as $\bar{t}(r) =t \, (\ll 1)$; then, the equations reduce to $\pd_r^2 \zeta +K^2 \zeta =0$ and $\pd_r^2 \theta - L^2 \theta = 0$, where $K= \sq{3(1 + a_1 t^2)} H_{\rm D} \simeq \sq{3}H_{\rm D}$ and $L = M_{\rm P} t/2 \, (\ll 1)$. From these, we obtain $\zeta \sim \sin (K r)$ and $\theta \sim \sinh (L r)$. Integrating these twice with $r$ and finding $f$ and $g$ satisfying the conditions (\ref{conditions at origin}) reveals that they behave like
\begin{eqnarray}
      f(r) \simeq c \sin (\sq{3}H_{\rm D} r) + d r
          \label{f near origin}
\end{eqnarray}
and $g \simeq c^\pp \sinh(Lr) + d^\pp r$ near the origin. Since $L \ll 1$, 
\begin{eqnarray}
     g(r) \simeq d^\pp r
        \label{g near origin}
\end{eqnarray}
is obtained after all. Here the coefficients $c, d, d^\pp$ cannot be determined unless the equations are completely solved by imposing the boundary conditions (\ref{conditions at edge}) at the edge. The behavior of $g$ shows that $\lap3 \Phi = \lap3 \Psi$ holds around the origin. 

\begin{figure}[h]
\begin{center}
\includegraphics[width=8cm]{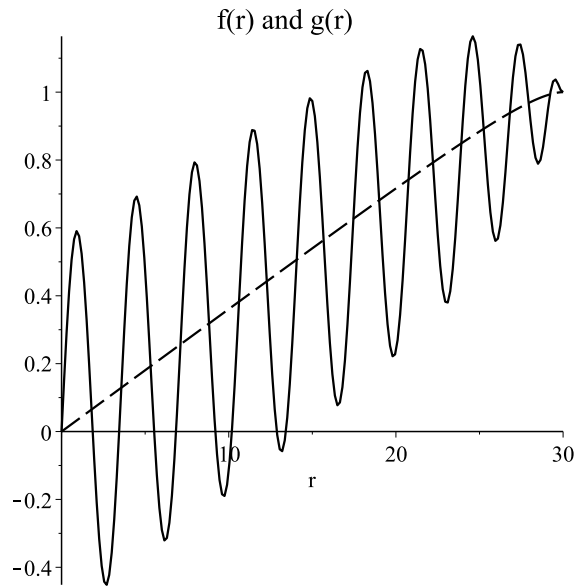}
\end{center}
\vspace{10mm}
\caption{\label{functions f and g}
Numerical results of $f$ (solid) and $g$ (dashed) in the case of $b_c = 10$, $\b_0 = 0.5$, $a_1 = 0.1$, and $H_{\rm D}/\Lam_{\rm QG}=60$. The unit is $H_{\rm D}=1$, then $m_{\rm pl}=1.784$ and $M_{\rm P}=0.356$.}
\end{figure}

\begin{figure}[h]
\begin{center}
\includegraphics[width=8cm]{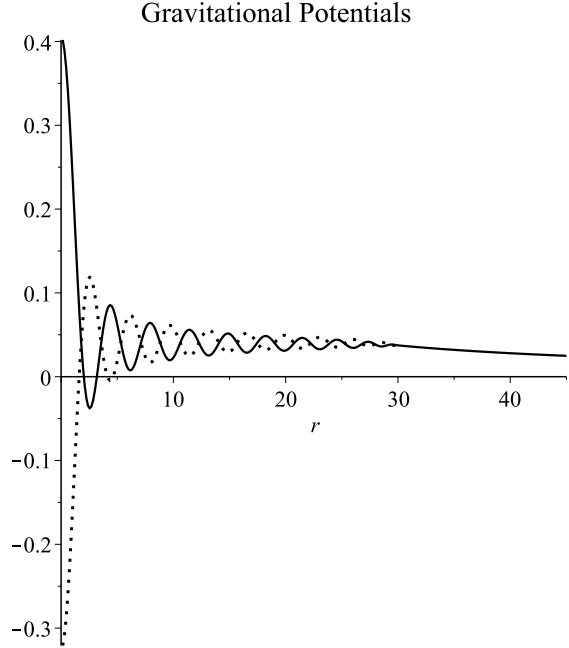}
\end{center}
\vspace{5mm}
\caption{\label{gravitational potentials}
The gravitational potentials $\Phi$ (solid) and $-\Psi$ (dotted) for an excited state with mass $m=2 m_{\rm pl}$. The tail $r \geq 30$ is the Schwarzschild solution $\Phi=-\Psi=r_g/2r$.}
\end{figure}

We numerically solve (\ref{equations of f and g}) as a boundary value problem with (\ref{conditions at origin}) and (\ref{conditions at edge}); then we obtain Fig. \ref{functions f and g}, where $H_{\rm D}/\Lam_{\rm QG}=60$, and $H_{\rm D}$ is normalized to be unity. The calculation is practically performed by setting the boundary condition at $r=R_h-\eps$ right inside the edge, and $\eps$ is brought close to zero until the result no longer changes. Here, $\eps=0.0001$. The gravitational potentials $\Phi$ and $-\Psi$ are then found as shown in Fig. \ref{gravitational potentials}. It can be seen that the behavior around the origin $(r \lesssim 5)$ of the numerical solution is $c = 0.543$, $d = 0.037$, and $d^\pp =0.036$.

The state mass is defined by $m = \int_{|\bx| \leq R_h} d^3 \bx \, T_{00}^{(4)} (\bx)$, where $T_{00}^{(4)}$ is the first two terms in (\ref{energy conservation}). Recall that the static solution satisfies (\ref{energy conservation}) of $\rho = 0$, thus rewriting the mass formula using this equation and $\Phi=(X+Y)/3=r_g(f+2g)/6r$ yields $m =  -2 M_{\rm P}^2 \int^{R_h}_0 4\pi r^2 d r \lap3 \Phi = (4\pi/3) M_{\rm P}^2 \, r_g \bigl[ f(R_h)+2g(R_h) \bigr] = 4\pi M_{\rm P}^2 r_g$, where the boundary conditions (\ref{conditions at origin}) and (\ref{conditions at edge}) are used. Thus, the relationship $r_g=m/4\pi M_{\rm P}^2 = 2Gm$ is derived.

\section{On time evolution}
Next, we examine how the static solution evolves with time. In this case, we have to solve the partial differential equations (\ref{trace equation}) and (\ref{constraint equation}), but unfortunately we cannot separate the coupled equations as we did when finding the static solution. Here, we will see the behavior around the origin where the coupling constant is small. Letting $t = 0$ and $B = 1$ and rewriting the equations with the variables being $X(\eta,r) = (r_g/2r) F(\eta,r)$ and $Y(\eta, r) = (r_g/r) G(\eta,r)$, we obtain
\begin{eqnarray}
   && ( \pd_\eta^2  - \pd_r^2 )^2 F(\eta, r) 
        + 2 \pd_\eta^2 ( \pd_\eta^2 -  \pd_r^2 ) G(\eta, r) 
                      \nonumber \\
   &&     - 3H_{\rm D}^2 \bigl[ ( \pd_\eta^2 - \pd_r^2 ) F(\eta,r) 
                                 + 2 \pd_\eta^2 G(\eta,r) \bigr] = 0 
        \label{partial equation for F} 
\end{eqnarray}
and
\begin{eqnarray}
   && (  3 \pd_\eta^2 - \pd_r^2 ) \, \pd_r^2 G(\eta,r) = 0 .
        \label{partial equation for G}
\end{eqnarray}

Let these equations be solved under the initial conditions $F(0, r)=f(r)$ and $G(0,r)=g(r)$, where $f$ and $g$ are the static solutions obtained above, and each behavior near the origin is given by (\ref{f near origin}) and (\ref{g near origin}). Since the gravitational potentials and each term of the energy equation (\ref{energy conservation}) do not diverge at the origin, the boundary conditions of $F(\eta, 0)=G(\eta, 0)=0$ and $\pd_r^2 F(\eta, 0)= \pd_r^2 G(\eta, 0)=0$ are imposed.

A solution of (\ref{partial equation for G}) allowed under these conditions is $G(\eta, r) = \tilde{d}(\eta) \, r$, where $\tilde{d}$ is an arbitrary function that satisfies the initial condition $\tilde{d}(0)=d^\pp$. We substitute this solution into (\ref{partial equation for F}) to find a solution of $F$. Further putting $F(\eta,r) = \hat{F} (\eta,r) - 2\tilde{d}(\eta) \, r$ gives $( \pd_\eta^2  - \pd_r^2 )( \pd_\eta^2  - \pd_r^2 - 3H_{\rm D}^2 ) \hat{F} (\eta,r) = 0$, where $\hat{F}$ also satisfies the same boundary conditions as $F$. From this, a general form of the solution satisfying the initial condition is $F(\eta,r) = \{ c + b \sin (\sq{3}H_{\rm D} \eta) + b^\pp [\cos (\sq{3}H_{\rm D} \eta) - 1] \} \sin (\sq{3}H_{\rm D} r) + \tilde{e}(\eta) r-2\tilde{d}(\eta)r$, where $b$ and $b^\pp$ are arbitrary constants, and $\tilde{e}(\eta)$ is a function satisfying $\pd_\eta^2 (\pd_\eta^2 -3H_{\rm D}^2)\tilde{e}(\eta)=0$ and $\tilde{e}(0) = d+2d^\pp$.

Now, $F$ and $G$ are not monotonic functions of time, except the terms involving $\tilde{d}$ and $\tilde{e}$. The behavior of these two terms cannot be determined by examining only near the origin. However, since all gravitational terms in (\ref{energy conservation}) that determines the matter energy density $\rho$ contain the spatial Laplacian, $\tilde{d}$ and $\tilde{e}$ do not affect changes in the energy density. In this way, we can see that $\rho$ does not monotonically increase at least near the origin. Thus, the excited state appears to be kept stable without the gravitational energy changing into matter.

\section{Conclusion and Discussion}

The dynamics of the asymptotically background-free quantum gravity begin to work at the energy scale $\Lam_{\rm QG}$ of $10^{17}$GeV below the Planck scale. This means that before reaching the Planck scale, the spacetime enters a new phase dominated by the conformal dynamics of the quantum gravity. The correlation length defined by the inverse of the scale is larger than the Planck length by 2 orders of magnitude, which gives the size of the excitation.

The existence of such excitations has long been predicted \cite{hamada02}, but it has not been able to construct it concretely due to the expected difficulty of nonlinearity. 
In this paper, although it is difficult to manage the nonlinearity in all mass ranges, we found that the linear approximation to the gravitational potentials can be successfully applied when the mass is several times the Planck mass. The mass is sufficiently larger than $\Lam_{\rm QG}$ so that it can be considered that the quantum gravity dynamics is activated inside, while the magnitude of the gravitational potentials is still small.

The progress was achieved by assuming the nonperturbative expression of higher-order corrections from the physical requirement that conformal dynamics disappear at the edge, and incorporating the effects of the running coupling constant approximated as a mean field that depends on the radial coordinate. The excitation was described as a solution of the linearized equations of motion for the gravitational potentials, and we found the solution given in Fig. \ref{gravitational potentials}. It can be regarded as a particle when viewed from the outside because its radius $R_h$ is larger than the horizon size $r_g$.

If the mass is smaller than $\Lam_{\rm QG}$, the quantum gravity will not be activated and no state will be excited. The coupling constant will remain large everywhere and the assumption that it is running will not be valid. On the other hand, a macroscopic object with a semiclassical horizon whose size is larger than $R_h$ looks like a black hole. It will undergo black hole evaporation and may eventually leave the small excited state as a remnant.

It is thought that many excited states were generated in the early Universe. Primordial black holes could be formed using them as seeds. If the state is actually stable or long-lived, it can be a candidate for dark matter as a purely gravitational object.

\end{document}